\documentclass[amsmath,amssymb,aps,prb, preprintnumbers,floatfix, twocolumn,showkeys,10pt]{revtex4-2}
\usepackage{graphicx}
\usepackage{dcolumn}
\usepackage{bm}
\usepackage{physics}
\usepackage[colorlinks=true, citecolor=red, urlcolor=blue]{hyperref}
\usepackage{url}
\usepackage{orcidlink}

\begin{document}
\title{Evaluating the transition dipole moment of quantum dots with absorption and angle resolved photoluminescence spectroscopy}
\author{Harshavardhan R. Kalluru\textsuperscript{1}\orcidlink{0000-0003-1639-0595}}
\email{kallurureddy@iisc.ac.in}
\affiliation{Department of Physics, Indian Institute of Science Bengaluru, 560012, India.\textsuperscript{\textup{1,2,3}}}
\author{Binita Tongbram\textsuperscript{2}\orcidlink{0000-0003-3443-837X}}
\affiliation{Department of Physics, Indian Institute of Science Bengaluru, 560012, India.\textsuperscript{\textup{1,2,3}}}
\author{Jaydeep K. Basu\textsuperscript{3}}
\affiliation{Department of Physics, Indian Institute of Science Bengaluru, 560012, India.\textsuperscript{\textup{1,2,3}}}
\date{\today}
\begin{abstract}
 In this manuscript, the evaluation procedure of transition dipole moment (TDM) is discussed. Semiconducting Cd\textsubscript{x}Zn\textsubscript{1-x}Se\textsubscript{y}S\textsubscript{1-y} alloyed quantum dots (AQDs) are used as the two level emitting system. The AQDs are then self-assembled into monolayers by the Langmuir-Schaefer method. The TDM magnitude and orientation of AQDs are extracted from the absorption spectrum and the angle resolved photoluminescence emission spectrum measurements respectively. The TDM of AQDs in vacuum is evaluated as 9.07 D and the anisotropy coefficient shows that the AQD emission is isotropic.
\end{abstract}
\keywords{Transition dipole moment, oscillator strength, spontaneous emission, photoluminescence, time resolved photoluminescence spectroscopy}
\maketitle
\section{Introduction}
Spontaneous emission process is at the heart of all light emitting devices (LEDs). Understanding the nature spontaneous emission rate enables design of strategies and protocols for optimizing the light emission efficiencies of LEDs. According to the semi-classical theory of light, under dipolar approximation, the spontaneous emission rate can be described in the terms of transition dipole moment ($\vec{\mu}$) and the photonic density of states ($\rho$). 

In this manuscript the dipolar approximation\cite{PhysRevA.97.032105} holds well, as the criteria for weak interaction between probing electromagnetic radiation and the emitting quantum dots is satisfied.
\begin{itemize}
\item The wavelength of light emitted ($\simeq$ 609 nm) is much larger than the size ($\simeq$ 6 nm) of the quantum dots.
\item The emission frequency of quantum dots ($\simeq$ 10\textsuperscript{14} s\textsuperscript{-1}) is significantly larger than the inverse of the emission lifetime ($\simeq$ 10\textsuperscript{8} s\textsuperscript{-1})
\end{itemize}
This dipolar approximation is a useful but primitive picture of quantum dot's spontaneous emission, as it ignores the Coulomb interaction of charge carriers, Hartree-Fock band gap renormalization and other carrier scattering processes.\cite{CHOW2013109} The effect of quantum dot confinement potential on the in{\-}plane motion of carriers is significant as the Bohr{-}Exciton radius of CdSe is 5.7 nm\cite{PhysRevLett.89.086801} is close to the size of the quantum dots. The quantum dot capping ligand molecules also influences its energy levels.\cite{10.1063/1.5128334}

Under the dipolar approximation, the Fermi golden rule, which indicates the probability\cite{RLoudon2000, scully1997} of a spontaneous emission rate ($\Gamma^\textup{vac}_{21}$) between two non-degenerate energy levels E\textsubscript{2} and E\textsubscript{1} of an emitter in vacuum is given as 
\[\Gamma^\textup{vac}_{21}=\frac{2\pi}{\hbar}|\vec{\mu}_{21}|^{2}{\rho}\tag{1}\]
The transition dipole moment (TDM) $\vec{\mu}_{21}$ of non-degenerate two level system\cite{ditchburn1976} is given by 
\[|\vec{\mu}_{21}|^{2}=e^{2}\cdot|\bra{1}r\ket{2}|^2\tag{2}\]
The TDM is directly related to the radiative decay lifetime and absorption cross-section of the two level system. Experimentally, it is convenient to measure radiative decay rate of a non-degenerate two level system in vacuum, which can be expressed in terms of the Einstein A and B coefficients,\cite{10.1119/1.12937} as shown in Fig. 1(a). The Einstein A and B coefficients describe the transition rate from the respective energy levels. A\textsubscript{21} is the spontaneous emission rate from energy level 2 to energy level 1. B\textsubscript{21} and B\textsubscript{12} and are the absorption and stimulated emission rates of level 2 and 1 respectively.

The Einstein A and B coefficients are inter-dependent. Under dipolar approximation, the Einstein A coefficient is the reciprocal of the spontaneous emission life time ($\tau^\textup{vac}_\textup{21}$) in vacuum. The Einstein A and B coefficients are related to the TDM.\cite{Stenzel2024,10.1119/1.12937,MCEHuber_1986}
\[\textup{A}^\textup{vac}_{21}=\frac{1}{\tau^\textup{vac}_\textup{21}}=\frac{2\omega^3_{21}}{3\epsilon_\textup{o}hc^3}\cdot|\vec{\mu}_{21}|^{2}\tag{3}\]

\[\textup{B}^\textup{vac}_{12}=\textup{B}^\textup{vac}_{21}=\frac{2\pi}{3\epsilon_{o}h^2}\cdot|\vec{\mu}_{21}|^{2}\tag{4}\]

Here $\omega_{21}=2{\pi}\nu_{21}$. $\nu_{21}$ is the frequency of transition from level 2 to level 1, c is the speed of light and $\hbar=\frac{h}{2\pi}$ is the normalized Planck's constant. 

To measure the TDM in vacuum, either A\textsubscript{21} or B\textsubscript{21} have to be extracted from respective lifetime or absorption cross-section of quantum dots in vacuum. The TDM magnitude can be extracted from either of the Einstein coefficients.

In practice a dilute solution of quantum dots in a known solvent is prepared. The absorption-cross section ($\sigma_\textup{abs}$) of the solution is measured and then the decay lifetime of quantum dots can be extracted from the measured value of $\sigma_\textup{abs}$.

 The absorption cross-section of quantum dots in solution is directly related to the lifetime of the quantum dots in solution as\cite{doi:10.1021/jp046127i}
 \[\textup{A}^{s}_\textup{21}=\frac{1}{\tau^{s}_{21}}=\frac{8\pi}{\lambda_\textup{em}^2}\cdot\int_{\nu_{i}}^{\nu_{f}}\sigma_{abs}(\nu)d\nu \tag{5}\]
 Here $\textup{A}_{21}^{s}$ is the decay rate and $\tau^{s}_\textup{21}$ is the lifetime of colloidal quantum dots in solution. The decay emission wavelength is $\lambda$\textsubscript{em}. The frequency limits of the absorption band are $\nu_{f}$ and $\nu_{i}$. 

 The Einstein B coefficient of quantum dots in solution is given by\cite{Wolfgang2013,MCEHuber_1986} 
 \[\textup{B}^\textup{s}_\textup{12}=\frac{\lambda_\textup{em}}{h}\cdot\int_{\nu_{i}}^{\nu_{f}}\sigma_{abs}(\nu)d\nu\tag{6}\]

  The AQD vacuum lifetime ($\tau_{21}^\textup{vac}$) can be extracted from the AQD lifetime in a solvent ($\tau_{21}^\textup{s}$) as given by\cite{Toptygin2003, PhysRevB.65.035327, C6NR09021D, PhysRevLett.81.1381}
  \[\tau_{21}^\textup{vac} = \frac{9\textup{n}_\textup{s}^{5}}{\Big(2\textup{n}^2_\textup{s} + 1\Big)^2}\cdot\tau_{21}^\textup{s}=\Big(\zeta^2\textup{n}_\textup{s}\Big)\tau_{21}^\textup{s} \tag{7}\]

   Here $\zeta$ is the local field correction factor and n\textsubscript{s} is the refractive index of solvent surrounding the quantum dot. The local field correction factor ($\zeta$) is derived under the real cavity model in a uniform dielectric medium.\cite{CJFB1973,CAO199715,PhysRevLett.74.880}
   \[\zeta= \frac{3\textup{n}_\textup{s}^{2}}{\Big(2\textup{n}^2_\textup{s} + 1\Big)}\tag{8}\]

   The spontaneous emission rate is driven by the ambient zero-point energy fluctuations.\cite{PhysRevA.97.032105} So the environment correction is an important factor to account for the evaluation of the Decay rate and the magnitude of the TDM in vacuum.\cite{PhysRevA.43.467} The lifetime in vacuum is used for calculating the TDM value in this manuscript.
    \begin{figure}[htp]
    \centering
    \includegraphics{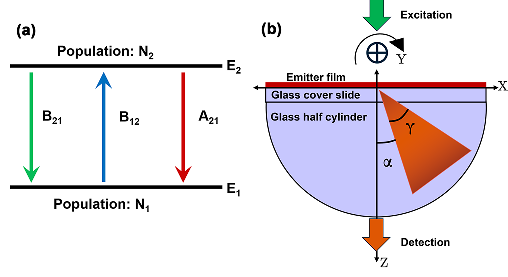}
    \caption{(a) shows the schematic of a two level system in equilibrium with radiation. (b) shows the emission cones of the light.}
    \label{fgr:1}
    \end{figure}
    To extract the orientation of TDM, the angular dependence of the emission has to be measured. This can be done by two techniques. One technique is the back-focal plane (BFP) imaging\cite{PhysRevApplied.14.034048} and the other is the angle resolved emission spectroscopy (ARPS).\cite{10.1364/FIO.2017.JTu2A.1,10.1063/1.5000478, 10.1063/1.3309705, 10.1039/C4TC00997E} 
    
    BFP imaging is the preferred method to find out the TDM orientation of single emitter. To sense the faint signal of single molecule, BFP imaging requires an advanced spectrometer system with large signal to noise (S/N) ratio. Typically a charge coupled device (CCD) detector cooled with liquid Nitrogen\cite{10.1063/1.4746251} or an Electron multiplying charge coupled device (EM-CCD) detector\cite{Backer2014} is used for the BFP imaging system, which makes it expensive. ARPS can be measured with a thermo\-electric cooled CCD, which is relatively affordable. ARPS is preferred for the ensemble measurement of TDMs of emitter films,\cite{VYSNIAUSKAS2022106655,doi:10.1021/acs.jpcc.3c06330} where signal is strong due to the large number of emitters in the spot.

    Typically an emitting film is deposited on a thin glass coverslip and attached to a half glass cylinder with refractive index matching oil. The emitting film is excited from the top and the emitted light is collected by the detector, through the half glass cylinder as shown in Fig.1(b). The substrate is set in XY plane. The light emission cone angle is $\gamma$ and it is off-set from the Z axis by an angle $\alpha$. 

    The emission spectrum anisotropy is quantified in the terms of emission anisotropy parameter ($a$),\cite{PhysRevApplied.14.064036} using emission angles $\alpha$ and $\gamma$ as
    \[a(\alpha,\gamma)=\frac{1}{3}\cdot\Big[cos^2\alpha + cos\alpha{\cdot}cos(\alpha+\gamma) + cos^2(\alpha+\gamma)\Big]\tag{9}\]
    Here $\alpha + \gamma{}\leq$ 90\textsuperscript{o}. The anisotropy coefficient ($a$) value range from 0 to 1. For an isotropic emitter the value of $a$ is 1/3. If an emission is completely oriented normal to the substrate plane, the value of $a$ is 1. For instance, Indium Selenide (InSe) layers have completely out of plane oriented excitonic dipole moment.\cite{Brotons-Gisbert2019,doi:10.1021/acs.nanolett.2c04902} If the value of $a$ is zero, then the light emission is completely oriented in the plane of the substrate. Molybdenum Sulfide (MoS\textsubscript{2}) monolayer is a good example for completely in the plane excitonic transition dipole moment.\cite{Schuller2013}

    The measurement of the emission intensity in the plane of substrate and the out of the plane of substrate, can give the value of the anisotropy coefficient. The measurement of one of the Einstein coefficients and the anisotropy coefficient gives the magnitude and direction of TDMs.

    \begin{table}
    \caption{The table shows the values of anisotropy parameter vs. TDM orientation, with the corresponding emission angles}
    \begin{center}
    \begin{tabular}{||c c c c||} 
     \hline
     TDM orientation & $\alpha$ & $\gamma$  & a\\ 
    \hline\hline
    Z & 0\textsuperscript{o} & 0\textsuperscript{o} & 1\\
    \hline
    XY & 90\textsuperscript{o} & 0\textsuperscript{o} & 0\\
    \hline
    Isotropic & 0\textsuperscript{o} & 90\textsuperscript{o} & 1/3\\
    \hline\hline
    \end{tabular}
    \end{center}
    \end{table}

    It is reported that display devices and LEDs have optimal emission efficiency and output optical power, when the TDMs of the emitters are oriented in the plane of substrate.\cite{PhysRevB.85.115205, PhysRevB.92.245306,10.1039/C4TC00997E,10.1038/s41467-021-22191-3} Due to the above mentioned considerations, determination of TDMs of emitters is needed in the LED and display industry. 

    Also from the perspective of controlling the interaction of emitters with a cavity, the knowledge of orientation is needed. If the cavity mode field ($\vec{\textup{E}}$) interacts with N emitters, the coupling coefficient (g)\cite{10.1088/0034-4885/78/1/013901,10.1038/nphys227,PhysRevApplied.18.014004} is given by
    \[\textup{g}=(\sqrt{\textup{N}})\vec{\mu}\cdot\vec{\textup{E}}\tag{10}\]
    By choosing optimally oriented TDMs ($\vec{\vec{\mu}}$) with appropriate concentration (N), the coupling (g) with cavity can be controlled.

    This manuscript describes the procedure for extraction of TDM magnitude and orientation of quantum dots. The mentioned procedure can be used for determination of TDMs of nanosized emitters as CdSe rings\cite{10.1038/s41467-019-11225-6} and CdSe platelets.\cite{10.1038/nnano.2017.177}

    \section{Methods}
    The AQDs used in this study are synthesized by one-pot hot injection method.(Appendix A) The synthesized AQDs are characterized by transmission electron microscopy (TEM), Energy Dispersive Spectroscopy (EDS) and atomic force microscopy (AFM) techniques as mentioned in Appendix B. The mean AQD size is extracted from AFM height profile as 5.6 nm $\pm$ 0.3 nm. TEM images are not considered for size estimation due to Argon plasma treatment done to AQDs before the TEM measurement.(Appendix B) The TEM and energy dispersive spectroscopy (EDS) measurements are performed with a Titan Themis 300 kV TEM system. The AQD quantum yield in Chloroform solvent is 0.33 $\pm$ 0.01 (Appendix C). The AQDs are self-assembled\cite{doi:10.1021/jp060416k,WO2005059952A2} in to films on glass and SiO\textsubscript{2}-Si substrates by the Langmuir Schaefer (LS) method.(Appendix D)

    The energy dispersive spectra of AQDs is measured for composition determination.(Appendix B: Table IV) The synthesized AQDs emission is in the visible region of light. So, the spontaneous emission process is Photoluminescence (PL). The steady state photoluminescence (PL) and time resolved photoluminescence (TRPL) are measured for the 5 transfer cycle (TC) sample using a confocal microscope (Witec alpha 300). 532 nm continuous wave (CW) laser and 405 nm pulsed lasers are focused onto the sample in reflection geometry using fiber coupled beam splitter and 20X magnification objective with numerical aperture (NA) 0.22 for PL and TRPL measurements respectively. 

    For PL measurement the 532 nm laser diode output power is set at 10 $\mu$W and the reflected light is filtered through a 532 nm edge notch filter to reject laser line. The filtered light is then relayed onto a 600 grooves/mm grating coupled Peltier cooled CCD detector. The grating and system configuration results in 0.13 nm spectral resolution. The CCD integration time is set at 5 s and averaged over 4 accumulation cycles. The PL intensity is measured in arbitrary units (a.u.).

    For TRPL measurement, the 405 nm pulsed laser is set at 40 $\mu$W power. The Laser pulse repetition rate is set at 20 MHz, which translates to 50 ns temporal separation between two consecutive pulses. The emitted light is collected with 20X/0.22 NA objective and filtered from the probing laser pulses using a 488 nm long pass filter. The integration time is 10 s and the emission is averaged over 10 accumulation cycles. The TRPL is measured by the time-correlated single-photon counting (TCSPC) method \cite{10.1016/B978-0-12-524140-3.50006-X} using a Picoquant SPAD detector with 30 ps lifetime resolution. The setup of angle resolved photoluminescence spectrometer is discussed in Appendix E.

\section{Results and Discussion}
\subsection{Absorption Cross-section and estimation of Lifetime of AQDs in solution}

    The Absorption cross-section ($\sigma_\textup{abs}$) of the AQD and Rh6G solutions is extracted from the absorbance (A) spectrum.\cite{C2JM30760J} The data is shown in Fig. 2.
    \[\sigma_\textup{abs}=\frac{\textup{ln10}\cdot\textup{A}}{\textup{N}_\textup{A}\cdot\textup{c}\cdot\textup{l}}\tag{11}\]
    Here N\textsubscript{A} is Avagadro number 6.022x10\textsuperscript{23}, c is the concentration and l is the path length of incident beam in cuvette.

    \begin{figure}[htp]
    \centering
    \includegraphics{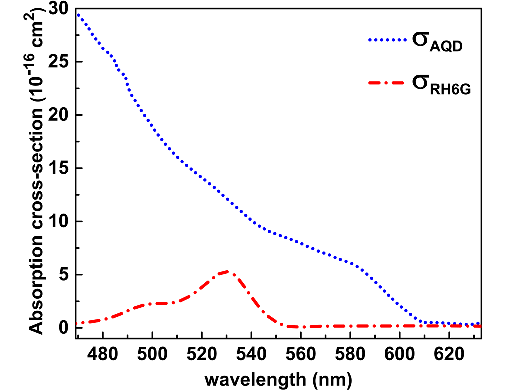}
    \caption{shows the absorption cross-section of the Rh6G dye and AQD in Chloroform solution}
     \label{fgr:2}
     \end{figure}
 
    The Rh6G peak absorption cross-section in Chloroform solvent is calculated as $\sigma_\textup{RH6G}$ = 5.29x10\textsuperscript{-16}${\ }$cm\textsuperscript{2} $\pm$ 0.07x10\textsuperscript{-16}${\ }$cm\textsuperscript{2}. The reported\cite{doi:10.1021/acsomega.7b00171} absorption cross-section of Rh6G peak absorption cross-section in water is $\sigma_\textup{RH6G}$ = 3.21x10\textsuperscript{-16}${\ }$cm\textsuperscript{2}. The difference in dye absorption cross-section is attributed due to the differing nature of the solvents.\cite{REISFELD1988142} The AQD peak absorption cross-section in Chloroform solvent is calculated as $\sigma_\textup{AQD}$ = 5.99x10\textsuperscript{-16}${\ }$cm\textsuperscript{2} $\pm$ 0.23x10\textsuperscript{-16}${\ }$cm\textsuperscript{2}.

    The intrinsic decay rate is $\textup{A}^{s}_{21}$ = 2.44x10\textsuperscript{8}${\ }$s\textsuperscript{-1} $\pm$ 0.09x10\textsuperscript{8}${\ }$s\textsuperscript{-1}. The corresponding solution lifetime of AQDs in Chloroform is $\tau^\textup{s}_\textup{21}$ = 4.10 ns $\pm$ 0.17 ns. This value extracted from the absorbance of a dilute AQD solution.The intrinsic value of the lifetime of AQDs in vacuum ($\tau^\textup{vac}_{21}$) is different from the lifetime obtained from the absorption cross-section of AQDs in solution. The measured Einstein coefficient B\textsubscript{21} = 3.29x10\textsuperscript{21}${\ }$ J\textsuperscript{-1}.m\textsuperscript{3}.s\textsuperscript{-2} $\pm$ 0.13x10\textsuperscript{21}${\ }$ J\textsuperscript{-1}.m\textsuperscript{3}.s\textsuperscript{-2}.

    Using the Chloroform refractive index n\textsubscript{s}=1.44,\cite{Kedenburg:12} the local field correction factor $\zeta$=1.21. The lifetime of AQDs in vacuum is $\tau^\textup{vac}_\textup{21}$ = 8.64 ns $\pm$ 0.36 ns. The corresponding Einstein coefficients for AQDs in vacuum are given by equations 24 and 25.
    \[\textup{A}^\textup{vac}_{21} = 0.12{\ }\textup{ns}^{-1} \pm 0.05{\ }\textup{ns}^{-1}\tag{12}\]
    \[\textup{B}_{12}^\textup{vac} = 4.94 \text{x}10^{20}{\ }\textup{J}^{-1}\textup{m}^3\textup{s}^{-2} \pm 0.21\text{x}10^{20}{\ }\textup{J}^{-1}\textup{m}^3\textup{s}^{-2}\tag{13}\]

    The TDM magnitude of AQDs in vacuum can be evaluated from equation (3) as follows
    \[|\vec{\mu}_{21}|^2=\frac{3\epsilon_{o}h{\lambda}^3}{16\pi^3}\cdot\frac{1}{\tau^{vac}_{21}}\tag{14}\]
    Generally the TDM values are reported in Debye units instead of SI units. Converting to debye units as, 1D = 3.335{x}10\textsuperscript{-30} C-m, the value of TDM turns out as $|\vec{\mu}_{21}|$= 9.07 D $\pm$ 0.19 D for AQDs in vacuum. The extracted TDM value of 9.07 D is in the range of the reported\cite{doi:10.1021/acsphotonics.7b00674,doi:10.1021/cm034081k, PhysRevB.79.045301} values for the quantum dots.

    The Einstein coefficient $A^\textup{vac}_{21}$ is estimated to be $0.12{\ }\textup{ns}^{-1} \pm 0.05{\ }\textup{ns}^{-1}$. The value of B\textsubscript{21} can be estimated from equation (4).(Appendix D)

\subsection{Steady State and Time Resolved Photoluminescence of AQD films}
 
    The steady state PL measurement is averaged over the objective focused spot. The objective spot size (d) is diffraction limited.\cite{Hecht2017} It is given by $\textup{d}=1.22\cdot\frac{\lambda}{\textup{NA}}$. The value of d for CW 532 nm laser is 2.95 $\mu$m and it is 2.25 $\mu$m  for 405 nm pulsed laser. So, the steady state PL spectra and TRPL measurements averages about 10\textsuperscript{5} AQDs. The measured spectra are shown in Fig. 3.

    The PL emission spectra is fitted with a Gaussian function and the peak position is at 608.75 nm $\pm$ 0.17 nm. The PL spectral line width is given by the Gaussian's full width half maximum (FWHM). The FWHM of 5 AQD layers is 47.83 nm $\pm$ 0.22 nm. 

    The TRPL decay curve is fitted with a single exponent decay function, to account the single decay channel available to the AQD transfer cycles i.e., decay into free space. The decay lifetime 5 AQD transfer cycles is extracted as 9.47 ns $\pm$ 0.06 ns. 

    The measured PL peak position ($\lambda_{21}$) is taken as 608.75 nm and the procedure for extracting Einstein coefficients in vacuum from the AQD solution absorption is discussed in Appendix D. The lifetime of AQDs in vacuum is extracted as $8.64 \textup{ns} \pm 0.36 \textup{ns}$. Considering the speed of light c =2.998x10\textsuperscript{8} m/s, the frequency ($\nu_{21}$) of the QD PL transition is given by
    \[\nu_{21}=\frac{2\pi}{\omega_{21}}=\frac{c}{\lambda_{21}}=4.9249\textup{x}10^{14}{\ }\textup{Hz} \pm 3\textup{x}10^{11}{\ }\textup{Hz}\tag{15}\]
    The TDM magnitude of AQDs in vacuum can be evaluated from equation (9) as follows
    
    The extracted value of $\tau^\textup{vac}_{21}$ = 8.64 ns is different from the experimentally measured value  $\tau^\textup{5TC}_{21}$ = 9.47 ns for the 5TC sample on SiO\textsubscript{2}-Silicon substrate. The lifetime of 5TC sample is within 10\% of the vacuum value ($\tau^\textup{vac}_{21}$). This indicates that the lifetime of AQDs does not vary significantly with the concentration of AQDs.

    \begin{figure}[htp]
    \centering
    \includegraphics{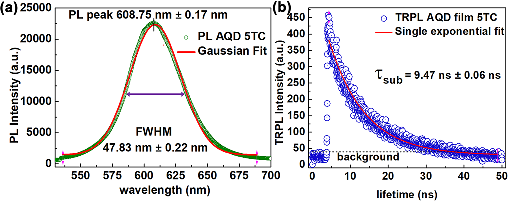}
    \caption{(a) shows the steady state PL emission spectra for 5 transfer cycles on silicon substrate. (b) shows the TRPL data for 5 transfer cycles on silicon substrate. The dashed line indicates dark count background (BG) of the SPAD detector. (c) shows the normalized absorbance data for Rh6G dye and AQDs. (d) shows the background corrected PL data for spin coated AQDs and Rh6G dye.}
    \label{fgr:3}
    \end{figure}

\subsection{Orientation of Transition Dipole Moments of Self-Assembled Alloyed Quantum Dot Films}

    The angle dependent PL emission spectra are measured by placing a linear polarizer plate in the beam path before the CCD input fiber. The polarizer (WP25M-VIS) has an average extinction rate of 800:1. The polarization axis of the linear polarizer can be rotated to align with horizontal and vertical directions. The vertical and horizontal axes are denoted by Y and X axes. The laser beam line is considered parallel to the Z axis.

    The far field intensity (I) of a point dipole is given by the magnitude of time averaged Poynting vector ($\vec{\textup{S}}$),\cite{10.1088/1367-2630/8/11/264,JDJackson} as 
    \[\Big<\vec{\textup{S}}\Big>=\textup{I}\hat{r}=\frac{\omega^4}{8{\pi}c^3}\frac{sin^2\theta}{r^2}\cdot|\vec{\mu}|^2\hat{r}\tag{16}\]

    \[\textup{I}\hat{r}=\frac{\omega^4}{8{\pi}c^3}\frac{sin^2\theta}{r^2}\cdot\Big(\vec{\mu}_{x}^2 + \vec{\mu}_{y}^2 +\vec{\mu}_{z}^2\Big)\hat{r}\tag{17}\]

    So, the intensity radiated by an ensemble of dipoles can be summed in terms of the intensities radiated by the individual dipole components along X, Y and Z axes.\[\textup{I}=\textup{I}_\textup{X} + \textup{I}_\textup{Y} + \textup{I}_\textup{Z}\tag{18}\]

    The substrate is assigned a co-ordinate system (X', Y' and Z') fixed with the laser beam spot center as origin. The angle dependence of polarized PL emission along X and Y axes is measured. The polarizer plate is removed and then the unpolarized angle dependent PL emission is measured. The polarized emission along Z axis can be evaluated by subtracting the net polarized PL emission from the unpolarized PL emission.

    The glass cylinder is rotated along the Y'=Y axis, which coincides with the symmetry axis of the glass cylinder. To extract the intensity from the TDMs along axes (X',Y' and Z') from the measured emitted intensity of TDMs along X, Y and Z axes. By ignoring the transmission losses, this can be done by a rotation of axes, by angle $\theta$ around Y axis.
    \[\begin{bmatrix}
    \textup{I}_{\textup{X'}}\\
    \textup{I}_{\textup{Z'}}\\
    \end{bmatrix}
    =
    \begin{bmatrix}
    cos\theta & sin\theta\\
    -sin\theta & cos\theta\\
    \end{bmatrix}
    \cdot
    \begin{bmatrix}
    \textup{I}_{\textup{X}}\\
    \textup{I}_{\textup{Z}}\\
    \end{bmatrix}
    \tag{19}\]
    \[\textup{I}_{\textup{Y'}}=\textup{I}_{\textup{Y}}\tag{20}\]
    Considering the substrate plane is defined as X'-Y' plane, the in plane ($||$) and out of plane ($\perp$) contributions can be expressed as
    \[\textup{I}_{||}=\textup{I}_\textup{X'}+\textup{I}_\textup{Y'}\tag{21}\]
    \[\textup{I}_{\perp}=\textup{I}_\textup{Z'}\tag{22}\]

    From the angle dependence of these three spectral measurements, TDM orientation can be estimated from the anisotropy coefficient as
    \[\textup{I}_\textup{o}=a\cdot\textup{I}_{\perp} + (1-a)\cdot\textup{I}_{||} \tag{23}\]

    \begin{figure}[htp]
    \centering
    \includegraphics{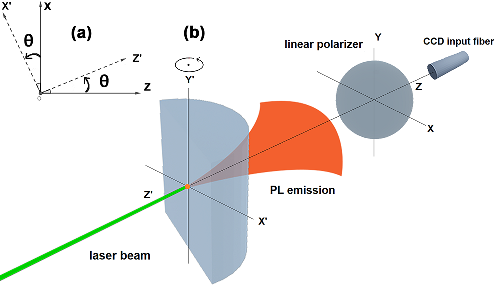}
     \caption{shows the schematic of ARPS measurement.}
      \label{fgr:4}
    \end{figure}

    The fitted spectral map for the data measured by ARPS setup is shown in Fig. 5(a) for 5TC sample. For 5TC sample, the full width half maximum (FWHM) of PL spectra is considered for further analysis. The polarization dependent PL spectra at each incident angle is obtained by integration over the FWHM. 

    The cylinder axis was deliberately misaligned with laser beam, by shifting the cylinder along X axis about 3 mm, in order to visualize the effect on angle resolved emission. The raw angle resolved emission pattern for misaligned and aligned system are compared in Fig. 5(b). For the misaligned system, the contrast between angles is poor and emission pattern is spread over the angles. In case of the aligned system, the emission pattern contrast is sharp. 

    The corresponding integrated spectral data is shown in Fig. 6(a) for 5TC sample.The data is only collected from 0\textsuperscript{o} to 80\textsuperscript{o}. For better visibility the data is duplicated till -80\textsuperscript{o}. The intensity contribution of in the plane (IP) and out of plane (OP) dipoles is extracted from the data using the equations (16-20).  The processed data is shown in Fig. 6(b), which encapsulates the far field emission pattern of ensemble of AQDs. Similarly the processed data is shown in Fig. 6(c) and Fig. 6 (d) for 3TC and SC samples. From the IP and OP emission patterns,the anisotropy parameter is extracted by using equation (14).

    \begin{figure}[htp]
    \centering
    \includegraphics{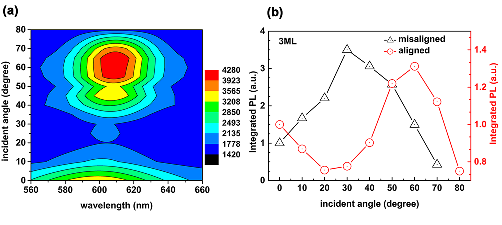}
     \caption{(a) shows the angle resolved unpolarized PL emission map of 5TC sample. (b) shows the variation in angle dependent unpolarized PL emission between the aligned and misaligned ARPS system for 3TC sample.}
    \label{fgr:5}
    \end{figure}

    A single CdSe-ZnS or CdSe quantum dot has a circular TDM due to the two fold degeneracy in the exciton states.\cite{PhysRevLett.93.107403,10.1073/pnas.0133507100}. In an ensemble, due to the averaging over 10\textsuperscript{11} dots, the TDM becomes isotropic. AQD ensemble PL emission is reported as isotropic.\cite{10.1088/1361-648X/abb650} The AQD 5TC sample also serves as a benchmark for the ARPS instrument.\\
    \begin{figure}[htp]
    \centering
    \includegraphics{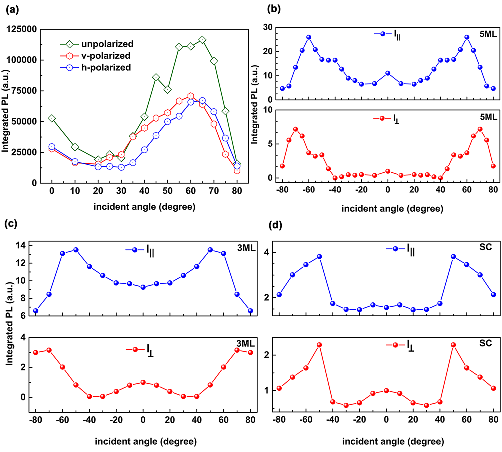}
     \caption{(a) shows the raw angle dependence of PL emission for 5TC sample.(b), (c) and (d) show the processed angle dependence of in-plane and out of the plane PL emission of 5 TC, 3TC and SC samples.}
    \label{fgr:6}
    \end{figure}
    The anisotropy coefficient represents the ratio of out of plane TDMs to the total TDMs in the beam spot. With the value of $a$, the orientation of TDM can be known. The mean anisotropy coefficient measured for the AQDs is mentioned in the table. II. The worst case anisotropy coefficient measured for 5TC AQD sample is 0.351, which amounts to a 6\% deviation from the expected value 0.333. The value agrees with the reported range of anisotropy coefficient for CdSe-CdS quantum dots is 0.33 to 0.35.\cite{PhysRevApplied.14.064036}

    The OP and IP contributions are extracted from the IP ($\textup{I}_{||}$) and OP ($\textup{I}_{\perp}$) intensities to the total intensity ($\textup{I}_{||} +\textup{I}_{\perp}$). The ARPS measurement affirms that the OP dipole contribution is minimal at normal incidence and it is significant at oblique incidence. So most of the light emitted at angles $<$ 40\textsuperscript{o} can be attributed to IP transition dipoles.
    \begin{table}
    \caption{The mean anisotropy coefficient measured for various AQD samples.}
    \begin{center}
    \begin{tabular}{||c c||} 
    \hline
    Sample & $a$\\[0.5ex] 
    \hline\hline
    SC & 0.430 $\pm$ 0.002\\ 
    \hline
    3TC &  0.291 $\pm$ 0.002\\
    \hline
    5TC &  0.349 $\pm$ 0.002\\
    \hline
    \end{tabular}
    \end{center}
    \end{table}

    The angular emission emission pattern changes significantly with the increasing AQD films, i.e. concentration of AQDs. The intensity contrast between near normal and oblique angles increases from SC to 3TC and is  largest for 5TC sample. The agreement of the anisotropy coefficient, with ideally expected value also improves in a similar trend.

    So this indicates that if the emitter quantum yield is lower than 33\%, then multiple layers of emitters are to be deposited on the substrate, to get an accurate angle resolved emission pattern. If the emitter quantum yield is near unity,\cite{10.1039/C2SC00561A} for example in the case of CdSe/CdS quantum dots, even spin-coated substrate can give an accurate angle resolved emission pattern.\cite{PhysRevApplied.14.064036}
    \begin{table}
    \caption{The table shows the angle dependent IP and OP contributions of total detected spectral intensity of various AQD samples.}
    \begin{center}
    \begin{tabular}{||c c c c||} 
    \hline
    Sample & angle & IP (\%) & OP (\%)\\[0.5ex] 
    \hline\hline
    SC & 0\textsuperscript{o} & 39 & 61 \\ 
    \hline
    SC & 70\textsuperscript{o} & 33 & 67 \\ 
    \hline
    3TC & 0\textsuperscript{o} & 90 & 10 \\
    \hline
    3TC & 70\textsuperscript{o} & 73 & 27 \\
    \hline
    5TC & 0\textsuperscript{o} & 92 & 8 \\
    \hline
    5TC & 70\textsuperscript{o} & 65 & 35 \\
    \hline
    \end{tabular}
    \end{center}
    \end{table}
    \begin{figure}[htp]
    \centering
    \includegraphics{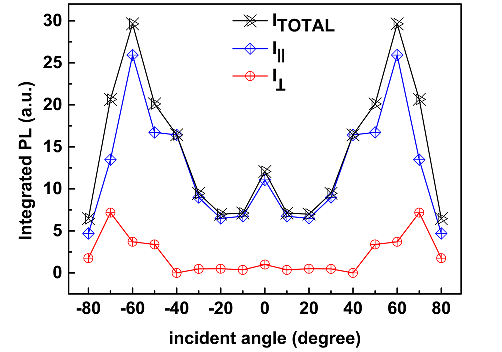}
     \caption{shows the angle dependence of in-plane and out of the plane PL emission relative to the total emission of the 5TC sample .}
     \label{fgr:7}
    \end{figure}
    The IP and OP contributions in Table III, from the SC sample deviate significantly from the thicker film samples (3TC and 5TC). This can be understood from the poor intensity contrast at normal incidence for SC sample. The agreement between 3TC and 5TC samples is good. Essentially the IP dipole emission is dominantly detected at normal incidence and the OP dipoles emission is dominantly detected at angles larger than 70\textsuperscript{o}. This is also shown in the Fig. 7. 
    \section{Conclusion}
    The synthesized AQDs are characterized by TEM, AFM and optical spectroscopy. The environment corrected TDM value of 9.07 D is obtained for the AQDs in vacuum. The PL emission of the ensemble is isotropic as the orientation of TDMs is distributed equally in all directions. The in the plane dipole contribution to PL emission is dominant at normal incidence and the out of plane dipole contribution to the PL emission is dominant at oblique angles of incidence.

\section{Author Contributions}
    HRK contributed to ideation, sample preparation, instrumentation, analysis and writing the manuscript. BT characterized the quantum dots with electron microscopy. JKB contributed to project management.
\begin{acknowledgments}
    Authors acknowledge the use of many mechanical components of the angle resolved emission spectroscopy (ARPS) setup from late Prof. Vasant Natarajan's lab, Physics department, Indian institute of science (IISc), Bengaluru. Authors acknowledge financial support from science \& engineering research board (SERB), India and Institute of eminence (IoE) grant from the Ministry of human resource development (MHRD) executed through IISc. B. Tongbram thanks the Department of science and technology (DST), Inspire faculty programme for fellowship. H.R. Kalluru thanks the Micro and nano characterization facility (MNCF-CeNSE), IISc for access to titan themis 300 kV TEM facility.
\end{acknowledgments}
\appendix
\section{Alloyed Quantum Dot Synthesis}
    For use as emitters in this study $\textup{Cd}_{x}\textup{Zn}_{1-x}\textup{Se}_{y}\textup{S}_{1-y}$ alloyed quantum dots (AQDs) are synthesized. The reported\cite{10.1007/s11144-014-0813-0, 10.1021/jp800063f} hot injection procedure is used for synthesizing AQDs. The glassware used for synthesis is cleaned in hot RCA SC-I solution\cite{10.1149/1.1390954} for 10 minutes. Then the glassware is cleaned in an ultrasound bath and subsequently rinsed thrice with ultra-pure DI water and dried in an oven.

    Cadmium oxide powder (CdO-99.5\%), Zinc Oxide powder (ZnO-99\%), Oleic Acid (OA-90\%), Selenium powder (Se-99.99\%), Sulphur powder (S-99.9\%) Trioctyl phosphine (TOP-99\%) n-Hexane (99\%) and Octadecene (ODE-99\%) are procured from Merck.

    25.68 mg CdO, 162.81 mg ZnO, 3.52 ml OA and 10 ml ODE are measured and dropped in a three-neck borosilicate glass flask of volume 50 ml, along with a Teflon magnetic stirrer bead. 21 mg Se powder, 84.6 mg S powder and 2 ml TOP is added in a borosilicate glass vial. The glass vial with Se-S-TOP precursors is flushed with Nitrogen gas and sonicated in an ultrasound bath for twenty minutes till the solution becomes completely transparent.

    The three-neck flask is placed on a magnetic stirrer plate in a heating mantle with a PID controller and temperature feedback sensor. The flask is then connected to a condenser column, which is connected to a Schlenk line, with a dual passage option for flushing the flask with gas and evacuating the flask. The flask necks are closed with rubber septa and sealed with Teflon tape to avoid air leakage into the flask. The flask is then evacuated with a turbo pump through the Schlenk line and the vacuum is maintained for 30 minutes, with the help of a vacuum pump. Then the PID is turned on with a set temperature $\textup{150}^{\textup{o}}$ C. The solution in flask is heated to the set temperature under vacuum. Then the flask is held at set temperature for 30 minutes under vacuum.

    Then the vacuum pump is turned off and the flask is flushed with Nitrogen gas and flow of nitrogen gas is maintained throughout the synthesis. As Nitrogen gas is heavier than air it settles in bottom half of the three-neck flask and forms a barrier against any atmospheric oxygen leaking in. The whole heating mantle is wrapped in glass-wool to insulate the flask and prevent loss of heat due to thermal radiation. The solution is flask should be pale yellow in colour. Now the set temperature of PID is changed to $\textup{310}^{\textup{o}}$ C. As soon as the temperature reaches $\textup{305}^{\textup{o}}$ C, 1.5 ml of Se-S-TOP solution is injected into the flask with help of a syringe and pipette.

    The reaction mixture rapidly turns orange and then dark red. The reaction is allowed to proceed for 15 minutes from injection time. The solution temperature is maintained at 305\textsuperscript{o} C $\pm$ 5 \textsuperscript{o} C for the duration. Then the PID is turned off and the flask is quenched in a water bath to stop the reaction. As the reaction mixture reaches room temperature, 15 ml of Acetone (99\%) and 5 ml Chloroform (99.9\%) is added to reaction mixture. The leftover precursors dissolve in the Acetone phase. The AQDs dissolve in the Chloroform phase and get separated from the precursors. This reaction mixture is further cleaned by centrifuging.

    After centrifuging for 10 minutes at 12000 rpm, the precipitate is collected and re-dispersed in 3 ml Chloroform. The supernatant solution is discarded in a glass bottle. Then 9 ml Acetone is added to Chloroform solution. This completes one cycle of cleaning. Two more such cycles are repeated. The precipitate is selected and the supernatant is discarded in every cycle. The precipitate after third cycle is dispersed in Chloroform and allowed to dry out in a desiccator. The AQD powder is stored in a cleaned glass vial in dark for further use.

\section{Structural composition of AQDs}
    The synthesized AQDs are characterized by transmission electron microscopy (TEM) and energy dispersive X-ray spectroscopy (EDS) measurements, by using a Thermofisher titan themis 300 kV TEM machine. 10 $\mu$g/ml AQD solution is prepared and is dropcasted on to a copper TEM grid and allowed to dry under ambient conditions. Then the grid is placed under vacuum for 24 hours. Before mounting onto TEM machine, the TEM grids are cleaned under Argon plasma for 40 seconds at 23 W power and 4.7x10\textsuperscript{-2} torr pressure to minimize the organic residues due to oleic acid ligands and hexane solvent. This is also essential to maintain vacuum, as the residual organic molecules on the TEM grid i.e., oleic acid and hexane, degass under electron beam. So Argon plasma cleaning changes the size of AQDs. The AFM imaging gives a more reliable mean value of the size of AQDs compared to TEM imaging.
    \begin{figure}[htp]
    \centering
    \includegraphics{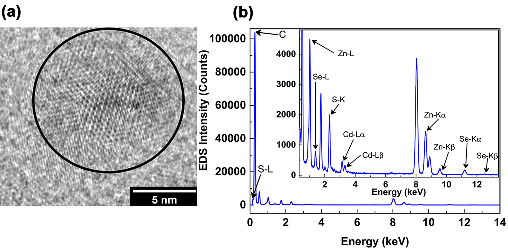}
    \caption{(a) shows the magnified TEM image of a single AQD. (b) shows the EDS spectra of AQDs. The inlay shows the magnified EDS spectra of AQDs.}
    \label{fgr:8}
    \end{figure}

    \begin{figure}[htp]
    \centering
    \includegraphics{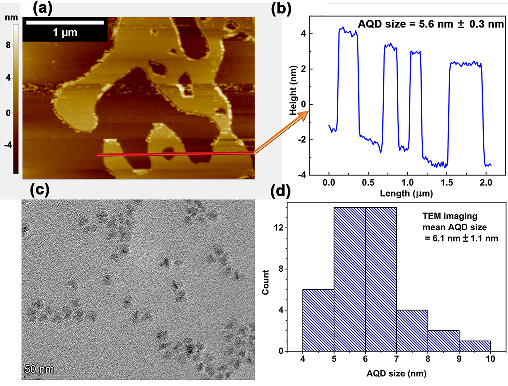}
    \caption{(a)shows NC-AFM image of the monolayer transferred after 3 compression cycles at 25 mm/min barrier compression speed and surface pressure of 35 mN$\cdot$m\textsuperscript{-1}. (b) shows the AFM height profile of the transferred monolayer.(c) shows the TEM image of AQDs and (d) shows the corresponding AQD size Histogram of AQD size}
    \label{fgr:9}
    \end{figure}

    The synthesized quantum dots are nearly spherical as shown in Fig. 8(a). EDS is a semi-quantitative tool used to analyze the chemical composition of alloys.\cite{10.1002/sca.21041,10.1016/0039-6028(89)90923-0} The EDS spectra of the AQDs have the characteristic X-rays of Cadmium, Selenium, Zinc and Sulphur as shown in Fig 8(b). The large carbon mass fraction is attributed to the organic ligand and solvent. The cationic and anionic atomic fractions of AQDs is extracted as x= 0.14 $\pm$ 0.1 and y= 0.13 $\pm$ 0.1. The EDS analysis concludes that AQD compositional formula is $\textup{Cd}_{0.14}\textup{Zn}_{0.86}\textup{Se}_{0.13}\textup{S}_{0.87}$
    \begin{table}
    \caption{The table shows the elemental analysis obtained from the EDS spectrum of AQDs.}
    \begin{center}
    \begin{tabular}{||c c c||} 
    \hline
    Element & Atomic fraction (\%) & Error (\%)\\ [0.5ex] 
    \hline\hline
    Carbon & 98.42 & 1.95\\ 
    \hline
    Sulphur & 0.66 & 0.12\\
    \hline
    Zinc & 0.7 & 0.09\\
    \hline
    Selenium & 0.10 & 0.01\\
    \hline
    Cadmium & 0.11 & 0.01\\
    \hline
    \end{tabular}
    \end{center}
    \end{table}

    The AFM image height profile in Figs. 9(a) and 9(b) indicates mean AQD size as 5.6 nm $\pm$ 0.3 nm, which is a better way to estimate AQD size. Plasma treatment prior to TEM analysis modifies the ligand on the quantum dot. So, AFM estimate represents more accurate size of quantum dots. The analysis of the TEM images in Figs. 9(c) and 9(d) indicate that the mean AQD size is 6.1 nm $\pm$ 1.1 nm. The size estimate is slightly less accurate relative to the AQD size estimate from AFM image. 

\section{Quantum Yield Determination of the AQD in solution}
    The PL quantum yield ($\phi$) of the AQDs is measured by the relative optical method.\cite{10.1021/ac900308v} In this method a fluorescent dye with known quantum yield is chosen as an standard sample. The emission and absorption spectrum of dye and quantum dots are measured. The quantum yield of QDs can be measured relative to the dye as
    \[{\phi}_{\textup{AQD}}={\phi}_{\textup{dye}}\cdot\frac{f_{\textup{AQD}}}{f_{\textup{dye}}}\cdot\frac{F_{\textup{dye}}}{F_{\textup{AQD}}}\tag{24}\]

    The Rhodamine 6g (Rh6G) dye is chosen as standard dye sample, as its quantum yield is known to be 0.75 in Chloroform solvent, at concentrations lower than 10\textsuperscript{-5} M, i.e. 12 $\mu$g ml\textsuperscript{-1}.\cite{10.1016/0009-2614(88)85073-5} 12 $\mu$g.ml\textsuperscript{-1} Rh6G solution is prepared in Chloroform solvent. Similarly AQD solution is prepared in Chloroform solvent with a concentration of 1 mg.ml\textsuperscript{-1}.

    Both the solutions are then spincoated on glass slides at 3000 rpm and for 60 s. The corresponding background corrected PL spectra are shown in Fig. 10. The spectra are integrated to calculate the spectral integrated PL factor $F$ given by
    \[F=\int{\textup{I}(\lambda)d\lambda}\tag{25}\]
    The integration is carried from 532 nm till the intensity reaches zero for AQDs and Rh6G. The absorption factor ($f$) is given by
    \[f=1-10^{-\alpha\textup{cl}}\tag{26}\]
    Here $\alpha$ is the specific Absorption coefficient, c is the concentration and l=1 cm is the length of cuvette used for measurement. 

    \begin{figure}[htp]
    \centering
    \includegraphics{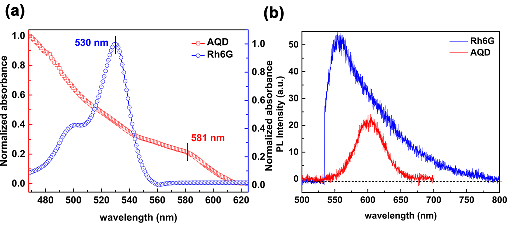}
    \caption{(a) shows the normalized Absorbance spectrum of Rh6G and AQDs at 12 $\mu$g.ml\textsuperscript{-1} and 0.675 mg.ml\textsuperscript{-1} concentration respectively. (b) shows the PL spectrum of AQDs and Rh6g}
    \label{fgr:10}
    \end{figure}

    The absorbance of Rh6G and AQD solutions is measured in a Perkin Elmer $\lambda$-35 UV-VIS spectrometer. The 12 $\mu$g.ml\textsuperscript{-1} Rh6g solution is made into two concentrations, 12 $\mu$g.ml\textsuperscript{-1} and 6 $\mu$g.ml\textsuperscript{-1}. The 1 mg.ml\textsuperscript{-1} AQD solution is diluted to two concentrations 0.675 mg.ml\textsuperscript{-1}, and 0.506 mg.ml\textsuperscript{-1}. Absorbance of solutions with two different concentrations of both Rh6G dye and AQDs are measured and the averaged absorption coefficient values are shown in Table V.
    \begin{table}
    \caption{The table shows the peak values of the averaged absorbance coefficient values of AQDs and Rh6G.}
    \begin{center}
    \begin{tabular}{||c c||} 
    \hline
    Emitter & Absorption coefficient ($\alpha$)\\[0.5ex] 
    \hline\hline
    AQD & 0.063 cm\textsuperscript{-1}.mg\textsuperscript{-1}.ml\\ 
    \hline
    Rh6g & 111.855 cm\textsuperscript{-1}.mg\textsuperscript{-1}.ml\\
    \hline
    \end{tabular}
    \end{center}
    \end{table}
    The spectral integrated PL $F$ factor for Rh6G and AQDs is 4245.32 a.u and 916.72 a.u. Plugging in the $F$ and $f$ factors in equation (11) and with $\phi_{\textup{dye}}$=0.75, the value of AQD quantum yield in the Chloroform solvent turns out to be 0.33 $\pm$ 0.01. The typical reported values of AQD quantum efficiency range from 0.1 to 0.8 depending on the reaction time.\cite{10.1021/cm070754d, Kim2022} The quantum yield varies with the solvent and dielectric environment of the synthesized quantum dots significantly.\cite{IBNAOUF2014369} AQDs are preferred over core-shell quantum dots, as the lattice mismatch in AQD is minimal due to composition gradient. The minimal lattice mismatch improves the quantum yield.\cite{10.1002/smll.200800841,IBNAOUF2014369}
\section{Self-assembly of AQDs into films}

    AQDs films are fabricated by self-assembly via Langmuir-Schaefer (LS) method. The LS method is a versatile method used to transfer films of colloidal quantum dots with hydrophobic ligands on to the substrates as Silicon, Glass and Quartz.\cite{10.1021/la904474h, 10.1021/acs.langmuir.6b01242} The LS setup consists of a Teflon trough coated with a hydrophobic coating. Similar hydrophobic Teflon barriers are placed on the trough and ultra{-}pure deionised (DI) water (resistivity-18.2 M$\Omega$) from a DI water system (Millipore) is filled in the trough till a water meniscus forms between the barriers. A pressure sensor is placed in between the barriers to sense the changes in surface pressure. 1.2 mg/ml AQD solution in n-Hexane is centrifuged for a fourth cycle, at 12000 rpm for 10 minutes. The supernatant solution is separated from the precipitate and is dispersed drop-wise using a Hamilton micro-syringe (50 $\mu$L capacity) between the barriers. Typically, 100 $\mu$L of AQD solution is used for a single transfer cycle.

    Once the AQD solution is dispersed, then the barriers are brought together using a stepper motor at a speed of 10 mm/min. This results in raising surface pressure, as the AQD are hydrophobic and compression of area between barriers results in mutual repulsive forces. If the AQDs are packed till monolayer limit, then further packing is not possible and as a result the surface pressure saturates. This indicates the formation of monolayer via self-assembly. At this point the substrate is lowered onto the area between barriers, by a motorized dipper at a rate of 10 mm/min. The substrate is allowed to touch the water surface and then is retracted at the same speed. The AQD film is now transferred onto the substrate and this whole transfer is a single cycle. The transfer is not always 100\% conformal and transferred monolayer is always patchy. The continuity and quality of the transferred layer, which varies largely with substrate, concentration of dispersed AQD solution, transfer speed.
 
    Any further transfer cycles run, will fill the gaps in transferred patchy monolayer. For instance, 3 transfer cycles will not result in 3 monolayers transferred. The thickness of transferred layers will be different from thickness 3 layers of AQDs and there will be a variation in thickness. As we are only interested in ensemble measurements, the samples are named after number of transfer cycles in this manuscript.
    \begin{figure}[htp]
    \centering
    \includegraphics{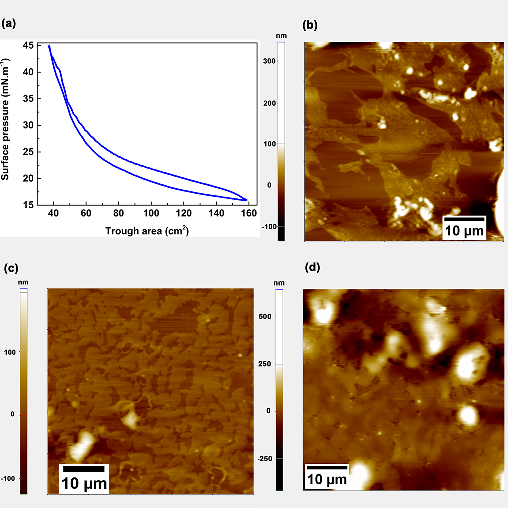}
    \caption{(a) shows a typical isothermal compression and expansion cycle of AQDs measured with our LS setup.(b), (c) and (d) show the NC-AFM images of 1 transfer cycle, 3 transfer cycles and 5 transfer cycles transferred after 10 compression cycles, at barrier speed 10 mm/min  and surface pressure of 42 mN$\cdot$m\textsuperscript{-1} respectively.}
    \label{fgr:11}
    \end{figure}
    The barrier compression speed and number of isothermal compression and relaxation cycles executed before the transfer determines the compactness of the AQD layer. Due to rapid barrier compression speed, large lateral monolayers cannot form, as compact self-assembly needs time for compact packing. Also multiple isotherms will seal the gaps between self-assembled monolayer domains. This has been verified by atomic force microscope (AFM) imaging of the monolayers in Fig. 11. The non-contact mode AFM is used for measurement, as it can measure precisely sub-nm range features of the sample.\cite{Hapala2015}

    Silicon wafers are ultrasonically cleaned in Acetone, then subsequently in Isopropyl alcohol and then dried in Nitrogen gas flow. These wafers are used as substrate for transfer of AQD layers and AFM imaging. An AQD monolayer is transferred onto Silicon substrate after 3 compression cycles at 25 mm/min barrier speed and surface pressure of 35 mN$\cdot$m\textsuperscript{-1}. NC-AFM images for the sample indicate that the monolayers are laterally small in size and are about 500 nm. 

    Compact AQD layer can be transferred on silicon substrates, only after 10 compression cycles and at barrier compression speed of 10 mm/min. Sequentially, 3 and 5 transfer cycles are sequentially transferred onto silicon wafers, after 10 compression cycles and at barrier compression speed of 10 mm/min and surface pressure of 42 mN$\cdot$m\textsuperscript{-1}. For all the three samples, the topography is imaged by AFM. The AFM imaging indicates that at least 10 cycles of compression are necessary for a compact monolayer, with large lateral dimensions. The substrate is wholly covered in case of 5 sequential transfer cycles. Considering CdSe/ZnS QDs aggregate in a hexagonal close packed structure at air-water interface,\cite{10.1016/j.ccr.2013.07.023} the maximum packing fraction is 74\%.\cite{AM1976} The area of 1 mm\textsuperscript{2} on the substrate, only 0.74 mm\textsuperscript{2} is filled by AQDs and each quantum dot is considered to be spherical in shape with 6 nm diameter. This translates to 2.48x10\textsuperscript{10} AQDs per transfer cycle per mm circular spot. For 5 transfer cycles, the density is 1.24x10\textsuperscript{11} per mm. So, the estimated order of emitters in a mm spot is 10\textsuperscript{11} for 5 transfer cycles.
\section{ARPS setup}
    ARPS system is assembled in our lab, for determining ensemble TDM orientation of emitters in transmission mode as shown in Fig. 12. The refined ARPS design\cite{PhysRevApplied.14.064036} was used for assembly of the spectroscopic system. The details of the optical components used in this setup are mentioned in table. VI.

    \begin{table}
    \caption{The table shows the details of the components used in the ARPS setup.}
    \begin{center}
    \begin{tabular}{||c c c c||} 
    \hline
    Marker & Catalogue number & Description & Manufacturer\\[0.5ex] 
    \hline\hline
    1 & CPS532 & DPSS laser & Thorlabs\\ 
    \hline
    2 & LB1676-ML & bi-convex lens & Thorlabs\\
    \hline
    N & ND06A/ND03A & ND filters & Thorlabs\\
    \hline
    3 & P1000K & pinhole & Thorlabs\\
    \hline
    4 & PR01 & rotation stage & Thorlabs\\
    \hline
    5 & AL2550M-A & aspheric lens & Thorlabs\\
    \hline
    6 & FELH0550 & long pass filter & Thorlabs\\
    \hline
    7 & WP25M-VIS & linear polarizer & Thorlabs\\
    \hline
    8 & PAF2S-11A & aspheric fiberport & Thorlabs\\
    \hline
    9 & QE-Pro & CCD & Ocean optics\\
    \hline
    F & QP600-2-UV-VIS & optical fiber & Ocean optics\\
    \hline
    \end{tabular}
    \end{center}
    \end{table}

    532 nm diode pumped solid state (DPSS) laser module rated at 4.5 mW is used for exciting Photoluminescence spectra from the AQDs. The laser beam path is aligned parallel to the optical table by using two 1 mm pin holes set at same height from the surface of optical table. Once the alignment is complete, the optical axis of the components is set by the laser beam path.

    \begin{figure}[htp]
    \centering
    \includegraphics{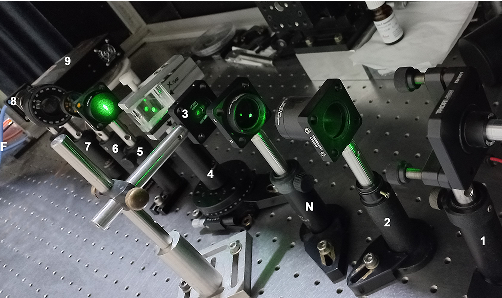}
    \caption{shows the ARPS instrument layout with the component markers. The markers are identified in table VI. The ND filters are placed in beam path, for photography purposes. For measurement, they are removed from the beam path.}
    \label{fgr:12}
    \end{figure}

    Then a bi-convex lens of focal length 100 mm is placed before the laser diode so that the pinhole and laser are on the both foci of the lens. Then a BK-7 glass half cylinder of radius 20 mm is placed on a rotating stage. The rotation axis of the stage is aligned with the cylinder symmetry axis. The laser beam spot size on the cylinder is about 3 mm. 

    An aspheric lens and an aspheric fiberport is set at 110 mm and 310 mm from the cylinder axis respectively. One end of an optical fiber is inserted inside the fiberport and the other end is inserted into a Peltier cooled CCD spectrometer input port. A 550 nm long pass filter and a linear polarizer is placed between the aspheric lens and fiber port.

    For ARPS measurement, three different concentrations have been selected. Three 18 mm x 18 mm glass coverslips are sonicated in Acetone and Isopropyl alcohol for 5 minutes and dried under nitrogen flow. 1 mg/ml solution of AQDs is prepared in n-hexane. 50 $\mu$L of the AQD solution is spin coated (SC) on to one cover slip for two cycles at 3000 rpm and 60s. 3 transfer cycles (3TC) and 5 transfer cycles (5TC) of AQDs are transferred onto coverslips by LS method. The samples are denoted by SC, 3ML and 5TC respectively. The cover-slide is stuck on a BK-7 half-glass cylinder (n\textsubscript{glass}=1.520) using a single drop of Leica index matching oil (n\textsubscript{oil}=1.518). Essentially the refractive index is uniform till the emitted light from the spot on the glass substrate reaches the cylindrical surface. Light does not undergo refraction at the cylindrical surface and the emission pattern of film is directly mapped out. The emission pattern is focused into the fiber port by the aspheric lenses. The 550 nm long pass filter, in the between the aspheric lenses filters out the laser line. 

    The linear polarizer plate is used to select the emission polarized along the along horizontal and vertical axes relative to the optical table plane, as shown in Fig. 3. As the light emitted by the dipoles is polarized along the TDM direction, the horizontal and vertical polarized emission intensity represents the dipoles oriented along the horizontal and vertical axes. By rotating the half cylinder, the polarized PL emission spectra of the AQDs deposited on the glass slide can be measured. The PL emission intensity of the AQDs measured without the polarizer plate in beam path, represent the emission due to all the dipoles in the incident laser beam spot. 

    From the angle dependence of these three spectral measurements, TDM orientation can be estimated from the anisotropy coefficient as
    \[\textup{I}_\textup{o}=a\cdot\textup{I}_{\perp} + (1-a)\cdot\textup{I}_{||}\]
    Here $a$ is the anisotropy parameter. $\textup{I}_{\perp}$ represents the intensity of dipoles normal to the substrate. $\textup{I}_{||}$ represents the intensity of dipoles parallel to the substrate. I\textsubscript{o} is the measured unpolarized intensity. To avoid the wavelength dependence, wavelength integrated PL spectra are used.
\bibliography{apssamp}
\bibliographystyle{apsrev4-2}
\end{document}